\documentclass[12pt]{entropy}

\usepackage{amsmath,amssymb,amsbsy,amsfonts}

\usepackage{graphicx}

\usepackage{url}

\title{The connection between entropy and the absorption spectra of 
Schwarzschild black holes for light and massless scalar fields.}

\author{Mendoza, S.$^{1,\star}$, Hernandez, X.$^{1,2}$, Rend\'on, P. L.$^{3}$,
Lopez-Monsalvo, C. S.$^{4}$ and Velasco-Segura, R.$^{3}$}

\address{$^1$Instituto de Astronom\'{\i}a, Universidad Nacional
                 Aut\'onoma de M\'exico, AP 70-264, Distrito Federal 04510,
	         M\'exico\\
         Email: sergio@astroscu.unam.mx. Email: xavier@astroscu.unam.mx\\
         $^2$GEPI, Observatoire de Paris, 5 Place Jules Janssen, 92195 Meudon, France.\\
         $^3$CCADET, Universidad Nacional
                 Aut\'onoma de M\'exico, Distrito Federal 04510,
	         M\'exico\\
         Email: pablo.rendon@ccadet.unam.mx. Email: rdroberto@gmail.com\\
         {$^4$School of Physics and Astronomy, University of
                 Southampton, Highfield, Southampton, SO17 1BJ, United
		 Kingdom\\
         Email: cesar.slm@gmail.com\\		 
$^{\star}$ Author to whom correspondence should be addressed.\\
{\em Received:  / Accepted:  / Published: }}
	}

\abstract{%
We present heuristic arguments suggesting that if EM waves with
wavelengths somewhat larger than the Schwarzschild radius of a black
hole were fully absorbed by it, the second law of thermodynamics would be
violated, under the Bekenstein interpretation of the area of a black hole
as a measure of its entropy.  Thus, entropy considerations make the well
known fact that large wavelengths are only marginally absorbed by black
holes, a natural consequence of thermodynamics.  We also study numerically
the ingoing radial propagation of a scalar field wave in a Schwarzschild
metric, relaxing the standard assumption which leads to the eikonal
equation, that the wave has zero spatial extent. We find that if these
waves have wavelengths larger that the Schwarzschild radius, they are
very substantially reflected, fully to numerical accuracy. Interestingly,
this critical wavelength approximately coincides with the one derived
from entropy considerations of the EM field, and is consistent with
well known limit results of scattering in the Schwarzschild metric.
The propagation speed is also calculated and seen to differ from the value
$c$, for wavelengths larger than $R_{s}$, in the vicinity of $R_{s}$. As
in all classical wave phenomena, whenever the wavelength is larger or
comparable to the physical size of elements in the system, in this case
changes in the metric, the zero extent 'particle' description fails,
and the wave nature becomes apparent.
}


\begin{document}

\section{Motivation}

 The presence of a black hole divides the universe into two causally
distinct regions.  Essentially, an event horizon hides a singularity
from the external universe with an inside region which is  causally
disconnected from the outside one.  The simplest example of a black hole
is the one given by the Schwarzschild metric.  These Schwarzschild black
holes have a ``spherical'' event horizon.  These holes are described
by a single parameter, their total collapsed mass \( M \) located at a
single point, in the ``centre'' of the event horizon.

 The problem of accounting for the apparent entropy decrease for the
universe when a body is swallowed by a black hole, hence making its
entropy disappear from the outside region, has been given an answer
\cite{Bek82} by associating an entropy to the area of the black hole
horizon.  This last quantity remains accessible to measurements performed
by external observers through its dependence on the black hole mass. The
extensive work on black hole thermodynamics of the 60s and 70s notably by
Bekenstein and Hawking (cf. \cite{nov} and references therein) has lead
to the establishment of the so called laws of black hole thermodynamics,
where black holes appear as classical thermodynamical objects, having
entropy $S_{BH}$ and temperature $T_{H}$ given by:

\begin{equation}
\label{SBek}
  \frac{ S_{BH} }{ k_{B} } = \frac{ A_{SH} }{ 4 A_{P} }
\end{equation}

and

\begin{equation}
\label{THawk}
  T_{H}= \frac{ hc }{ 8 \pi^{2} k_{B} R_{s} }.
\end{equation}

  In the above equations $A_{SH}= 4 \pi R_{s}^{2}$ is the area of the
Schwarzschild event horizon, $ A_{P} = \hbar G /c^{3}$ is the Planck
area and $R_{s}=(2 G M) /c^{2} $, the  Schwarzschild radius of a black
hole of mass M.  It can be seen, for example, that the merger of two
black holes of equal mass will result in a net increase in entropy, and
hence an event one should expect could happen, in terms of the second
law of thermodynamics.

  We now point to a particular process which would appear to violate the
second law.  Suppose a black hole of mass $M$ is absorbing an amount of
classical black body radiation having an energy $E_{EM}$, temperature
$T_{EM}$ and entropy $S_{EM}$.  When far from the black hole, this
entropy is given by

\begin{equation}
  \frac {S_{EM}} {k_{B}} = \frac {4}{3} \frac {E_{EM}}{k_{B}T_{EM}}.
\end{equation}

  The total entropy before the radiation is swallowed by the black hole
will therefore be $S_{BH} + S_{EM}$. After swallowing the radiation, the
black hole will experience an increase in mass $\Delta M=E_{EM}/c^{2}$,
and hence an increase in entropy given by equation (\ref{SBek}) as:

\begin{equation}
\label{DelSBek}
\frac{\Delta S_{BH}}{k_{B}} = \frac{8 \pi^{2} G}{h c} M \Delta M,
\end{equation}

\noindent in this case,

\begin{equation}
\frac{\Delta S_{BH}}{k_{B}} = \frac{8 \pi^{2} G}{h c^{3}} M E_{EM}.
\end{equation}

  We can now write the quotient of this increase to the original black
body radiation entropy, which was lost to the universe on it being
swallowed, as:

\begin{equation}
\frac{\Delta S_{BH}}{S_{EM}}=\frac{6 \pi^{2} G} {h c^{3}} k_{B} T_{EM} M. 
\end{equation}

  Now using equation (\ref{THawk}) we can write,

\begin{equation}
\label{SChange}
\frac{\Delta S_{BH}}{S_{EM}}=\frac{3 \pi}{8} \frac{T_{EM}}{T_{H}}.
\end{equation}

The right hand side of the above equation becomes $<1$ for
$T_{EM}<8T_{H}/3 \pi$.  We see that the process of a black hole swallowing
Planck radiation of a temperature somewhat lower than $T_{H}$ results
in an overall decrease in the entropy of the universe. Seen in this way,
this process would violate the second law of thermodynamics.  It would
appear to follow that at least part of the impinging radiation should not
be absorbed.  The following considerations point to which part of this
radiation one might expect not to be absorbed.

We now consider a black hole of mass $M$ swallowing a single
photon of wavelength $\lambda_{\gamma}$ and caring an energy
$E_{\gamma}=hc/\lambda_{\gamma}$. The increase in the black hole mass
will now be $\Delta M=h/c \lambda_{\gamma}$, to which there corresponds
an increase in the area of the Schwarzschild horizon of:

\begin{equation}
  \Delta A =8 \pi R_{S} \frac{h G}{c^{3} \lambda_{\gamma}},
\end{equation}

\noindent which we can write as:

\begin{equation}
\label{AreaQ}
\frac {\Delta A}{A_{P}} =16 \pi^{2} \frac{R_{S}}{\lambda_{\gamma}}.
\end{equation}

  We see that the increase in the black hole area becomes arbitrarily small,
in particular less than the Planck area, for photons having wavelengths
somewhat larger than the Schwarzschild radius of the black hole swallowing
them. This last might seem uncomfortable if one adopts the point of view
that the smallest dimension of area which should appear in any physical
theory or process is the Planck area, e.g.
a loop quantum gravity approach.  The case of a ``single photon'' is of
course an idealisation which strictly should be treated in the quantum
regime. For a black hole in the quantum regime itself, the results of
\citep{unruh} already imply no violation of the second law of
thermodynamics, it is the macroscopic limit what we will consider here.

The preceding two thought experiments lead to the conclusion that
either equation (\ref{SBek}) is not valid and arbitrarily small
area increases are allowed for black holes, or radiation colder than
$T_{H}$ (or photons with wavelengths longer than $R_{S}$) can not be
fully swallowed by black holes, if one wants to keep the second law of
thermodynamics.  We see that entropy considerations dictate the outcome of
physical processes around black holes.  The detailed physical mechanism
involved has to be explored in the scope of scattering problems in the
Schwarzschild space--time, as it has been done in the past by several
authors \citep{regge,vishveshwara,scattering,fabbri}, however entropy
constrains offer qualitative indications of the results to be expected.
The results of \citep{fabbri} already point to an upper limit  for
the wavelength of an EM wave above which absorption by a black hole
strongly decreases, of order \( R_S \).  We note that past studies of the
interaction between waves and black holes which identify upper critical
wavelengths, typically fall within the scope of WKB approximations,
where monochromatic waves are assumed to remain as such.  As will be
seen in the following, the particular nature of the problem makes the
above assumption invalid.  Indeed, as already noted by \citep{fabbri},
the error on the transmission coefficients under such approaches is of
order \( M / \lambda \), somewhat worrisome as the critical wavelengths
are typically of the order \( 1 / M \).

 The suggestion that EM wavelengths longer than the Schwarzschild
radius cannot be absorbed by a black hole is interesting, since going back
to equation (\ref{AreaQ}) implies that the smallest increase in the area
of a black hole which can result from the absorption of a photon, will
be of order $16\pi^{2} A_{p}$. Bearing in mind that whereas the quanta of
action $\hbar$ can be inferred from experiments due to its direct effect
on observables of the electro-magnetic field, the quanta of area cannot,
and has only been estimated to lie close to $A_{p}$ on dimensional
grounds. The above results might hint at the quanta of area $A_{q}$,
being of the order of $16\pi^{2} A_{p}$, with the quanta of length
resulting an order of magnitude larger than the canonical Planck length.

  We now turn to the propagation of EM radiation in the vicinity of a
black hole, to see if any mechanism to prevent radiation with wavelengths
larger than $R_{S}$ from being swallowed might naturally arise.  
The full EM problem, even with some simplifications, leads to
complicated, coupled differential equations. As a first approach we
simplify to an equation which is valid for the problem of
the interaction of a scalar field and a black hole. 

\section{Scalar waves in the Schwarzschild space--time}

  We start this section with the problem of a propagating EM wave,
moving towards a Schwarzschild black hole.  On first impression one
would naively jump to the conclusion that as photons can be treated
as massless particles, and hence, any form of EM radiation, composed
of photons, should simply follow null geodesics into the black hole.
However, such a treatment is derived under the assumption of vanishingly
small wavelengths.  See for example \cite{Land,Tol}, where the geometrical
optics approximation for EM waves is derived by requiring that the metric
can be treated as locally flat over the spatial variations in the studied
wave. The eikonal equation, trajectories described by $ds=0$ are, strictly
speaking, approximations to the propagation of EM radiation valid only
when the wavelengths can be treated as zero. In general, whenever the
dimensions of a system are much larger than the wavelength of radiation
moving around in it, light can accurately be treated as point particles
having zero extent. However, whenever elements appear with dimensions
comparable or smaller than the wavelengths of the radiation present,
the wave nature of light is immediately apparent, and EM radiation
must be treated explicitly as a wave. In the case of radiation having
wavelengths comparable or larger than the Schwarzschild radius of a black
hole, it is the variations in the metric which become comparable to the
wavelength of the EM waves. We must therefore treat the problem outside
of the eikonal approximation, and study the full physics of it.

  As it will be shown below, the problem becomes highly complex near \(
R \sim 1 \), with even simple wave pulses developing complex structures in
frequency space and exhibiting a large range of propagation speeds.  Thus,
a tortoise coordinate system where the characteristics of the equations we
are trying to solve correspond to waves with propagation speeds equal to
that of light, will result cumbersome and impractical.  The coordinate
singularity at the event horizon in the Schwarzschild space--time will not
be of concern as we are only interested  in a careful treatment of the
reflection of long wavelength waves which occurs before they reach the
event horizon.  Examples of studies treating wave reflection of black holes
in the Schwarzschild metric can be found in \citep{andersson,crispino,
scattering}.  The well known absorption of short wavelength waves will
not be treated in this article. 

The electromagnetic potential one--form \( \mathbf{A} \) is related
to the  Faraday 2--form \( \boldsymbol{F} \) by \citep{wheeler} \(
\boldsymbol{F} := \boldsymbol{\mathrm{d}} \boldsymbol{A} \), where \(
\boldsymbol{\mathrm{d}} \) represents the exterior derivative.  In the
absence of charge--currents, one pair of Maxwell's equations are given by
\( \boldsymbol{\delta} F = \boldsymbol{\delta\,\mathrm{d}} \boldsymbol{A}
= 0 \), where \( \boldsymbol{\delta} \) represents the codifferential
operator.  With this, the electromagnetic potential one--form satisfies
the relation \( \Delta \boldsymbol{A} - \boldsymbol{\mathrm{d} \, \delta}
\boldsymbol{A} = 0 \), where \( \Delta = \left( \boldsymbol{\mathrm{d}} +
\boldsymbol{\delta} \right)^2 \) represents Laplace--de~Rham's operator.
By imposing the Lorenz gauge given by \( \boldsymbol{\delta A } = 0 \)
it then follows that the electromagnetic potential one--form satisfies
a wave--like equation given by \citep{friedrich} \( \Delta \mathbf{A}
= 0 \).  In components, this equation yields:

\begin{equation}
  \begin{split}
  A^{\alpha;\beta}{}{}_{;\beta} = & \frac{ 1 }{ \sqrt{ -g } } \frac{ \partial
    }{ \partial x^\beta } \left\{ \sqrt{-g} \frac{ \partial A^\alpha }{
    \partial x_\beta } \right\}  \\
    & + 2 \Gamma^{\alpha}{}_{ \lambda \beta } \frac{
    \partial A^\lambda }{ \partial x_\beta } + \frac{ A^\theta
    \Gamma^\alpha{}_{ \theta \rho } }{ \sqrt{-g} } \frac{ \partial }{
    \partial x^\mu } \left( \sqrt{-g} g^{\mu\rho} \right) \\
    & + g^{\beta\rho} A^\theta \frac{ \partial \Gamma^\alpha{}_{\theta\rho} }{
    \partial x^\beta } + g^{\beta\rho} \Gamma^\alpha{}_{\lambda\beta} 
    \Gamma^{\lambda}{}_{\theta\rho} A^\theta = 0.
  \end{split}
\label{tthe-equation}
\end{equation}

  Limit and approximate solutions of this problem exist in the literature
\citep{scattering}.  In what follows, we include a simplified presentation
leading to a full numerical integration.  If we retain only the first
term of the above sum, for a particular component \( \Psi \) of \(
A^\mu \) this results in \citep[see for example][]{andersson}:

\begin{equation}
  \frac{ 1 }{ \sqrt{ -g } } \frac{ \partial
    }{ \partial x^\beta } \left\{ \sqrt{-g} \frac{ \partial \Psi }{
    \partial x_\beta } \right\} = 0.
\label{the-equation}
\end{equation}

\noindent  We note that this equation rigorously represents
the propagation of a massless scalar field \( \Psi \) in a curved
space--time, which is given by \( \Delta \Psi = 0 \) \citep{scattering}.
The solution of equation~\eqref{the-equation} will shed some light
as to the qualitative behaviour of the full EM solution, e.g. \citep{andersson}
studies the reflection of scalar waves from
black holes under a spectral decomposition analysis, as a first
order qualitative model for EM and gravitational waves.  We shall now
concentrate on solving equation \eqref{the-equation}, rigorously valid
for a scalar field in vacuum.

Taking a $(-,+,+,+)$ signature the Schwarzschild metric becomes

\begin{equation}
ds^{2}=\left(-1+\frac{2 M}{r}\right) dt^{2} +\left(1-\frac{2 M}{r}\right)^{-1} dr^{2}+r^{2}d\Omega^{2},
\end{equation}

where

\begin{equation}
d\Omega^{2}=d\theta^{2} + sin^{2} \theta d\phi^{2}.
\end{equation}

resulting in:

\begin{equation}
g_{\mu \nu}=\text{diag}\left[\left(-1+\frac{2 M}{r}\right), \left(1-\frac{2 M}{r}\right)^{-1}, r^{2}, 
r^{2}sin^{2}\theta \right],
\end{equation}

\begin{equation}
\sqrt {-g}=r^{2} sin\theta
\end{equation}

and,

\begin{equation}
g^{\mu \nu}=\text{diag}\left[\left(-1+\frac{2 M}{r}\right)^{-1}, \left(1-\frac{2 M}{r}\right), r^{-2}, 
\frac{1}{r^{2}sin^{2}\theta} \right]
\end{equation}

  We will retain the approximation that the metric is not modified by
the presence of the scalar field, vanishing field
strength, but otherwise introduce no approximations on the derivatives
of $\Psi$. We will not introduce a vanishing wave dimension.  For the
Schwarzschild metric, equation (\ref{the-equation}) reduces to:

\begin{equation}
\label{GREM}
  \bigg\{ \left(-1+\frac{2 M}{r}\right)^{-1} \frac{\partial^{2}}{\partial
   t^{2}} + 
    \bigg[ \frac{2}{r} \left(1-\frac{2
   M}{r}\right)\frac{\partial}{\partial r} + \frac{ 2 M }{r^2}
   \frac{\partial}{\partial r}
   + \left(1-\frac{2 M}{r} \right) \frac{\partial^{2}}{\partial r^{2}}
    \bigg] \bigg\} \Psi=0,
\end{equation}

\noindent where we have dismissed all non-radial spatial derivatives,
as we are interested only on a purely radially propagating scalar
wave. Further algebra reduces the above equation to:

\begin{equation}
\label{GREM2}
  \frac{1}{c^{2}} \frac{\partial^{2} \Psi}{\partial
    t^{2}}=\frac{(r-R_{S})^{2}}{r^{2}}\left(  \frac{\partial^{2}
    \Psi}{\partial r^{2}} + \frac{(2r-R_{S})}{r(r-R_{S})} \frac{\partial
    \Psi}{\partial r} \right).
\end{equation}

  It is clear that for $r>>R_{S}$ the above equation reduces to the
classical spherical wave propagation equation, as should be expected,
since the Schwarzschild metric reduces to Minkowsky space--time for
$r>>R_{S}$.

  Defining dimensionless quantities for the problem, $T:=c t/R_{S}$ and
$R:=r/R_{S}$, equation (\ref{GREM2}) becomes:

\begin{equation}
\label{DimLess}
  \frac{\partial^{2} \Psi}{\partial T^{2}}= \left( \frac{R-1}{R}
   \right)^{2} \frac{\partial^{2} \Psi}{\partial R^{2}}
   +\frac{(2R-1)(R-1)}{R^{3}} \frac{\partial \Psi}{\partial R}.
\end{equation}

If we now propose a solution of the form $\Psi(R,T)=F(T) G(R)$, we can
separate equation (\ref{DimLess}) to obtain:

\begin{equation}
\label{timeDless}
  \frac{d^{2} F}{d T^{2}} =- W^{2} F
\end{equation}

\noindent and

\begin{equation}
\label{RDless}
  \frac{(R-1)^{2}}{R^{2}} \left( \frac{d^{2} G}{d R^{2}}  
  +\frac{(2R-1)}{R(R-1)} \frac{d G}{d R} \right)= -W^{2} G
\end{equation}

Where  $W=\omega R_{S}/c$ and the separation constant was chosen $<0$
to guarantee periodic wave propagation. The solution to equation
(\ref{timeDless}) is trivial, while equation (\ref{RDless}) requires
somewhat more treatment.

We propose an approximate solution of the form $G(R)=U(R) \text{exp}(i
K R)$, which is then introduced into  (\ref{RDless}), the imaginary
component of which yields:

\begin{equation}
2 R \frac{dU}{dR} + \left( \frac{2R-1}{R-1} \right) U=0,
\end{equation}

\noindent and so,

\begin{equation}
U=\frac{C}{R^{1/2}(R-1)^{1/2}}.
\end{equation}

  Again, it is clear that for $R>>1$ the geometric dilution factor of $1/R$
on the potential, and hence $1/R^{2}$ on the energy, for spherical wave
propagation is recovered.  The real part of equation (\ref{RDless})
after introducing the trial solution gives the dispersion relation for
the problem:

\begin{equation}
\label{Disp}
K^{2}=\frac{1+4W^{2}R^{4} }{4R^{2}(R-1)^{2}}.
\end{equation}

  We can see that the standard $K=W$ dispersion relation is recovered for
$R>>1, W>>1$. At this point we can calculate the propagation velocity
as $W/K$,

\begin{equation}
\label{Vel}
  V=\frac{2WR(R-1)}{(1+4W^{2}R^{4})^{1/2}}.
\end{equation}

  We obtain the surprising result that the radial propagation of the
scalar wave towards a black hole does not always proceed at speed $c$
(recall that $c=1$ in the above units), but actually slows down on
approaching $R_{S}$, at a rate that depends on the frequency of the
wave, the effect becoming increasingly strong as $W$ decreases. Part
of the above effect, that at $W>>1$ is simply a consequence of the
Schwarzschild coordinate system, in which an inertial observer calculates
infinite travel times for waves travelling towards a black hole, however,
a chromatic effect is introduced by having considered a fuller physics
than the standard zero wavelength approximation. Again, in the limit
$W>>1, R>>1$ we recover $V=1$.

  We have obtained a dispersion relation which shows clear deviations from
the standard expression for the standard \( \mathrm{d}s=0 \) solution
corresponding to $W \sim 1$ and smaller. We notice that the above
analysis offers only a local approximation, a first correction valid
for non-zero, but small wavelengths, for which $K$ can be considered
approximately constant; a fuller solution to the problem requires a
numerical treatment of equation (\ref{DimLess}).  The dependence of \( K \)
with \( R \) exemplifies the limitations of any spectral WKB
approach to the problem.

\section{Numerical calculations}

\begin{figure}
\begin{center}
\includegraphics[height=8.5cm,width=9.3cm]{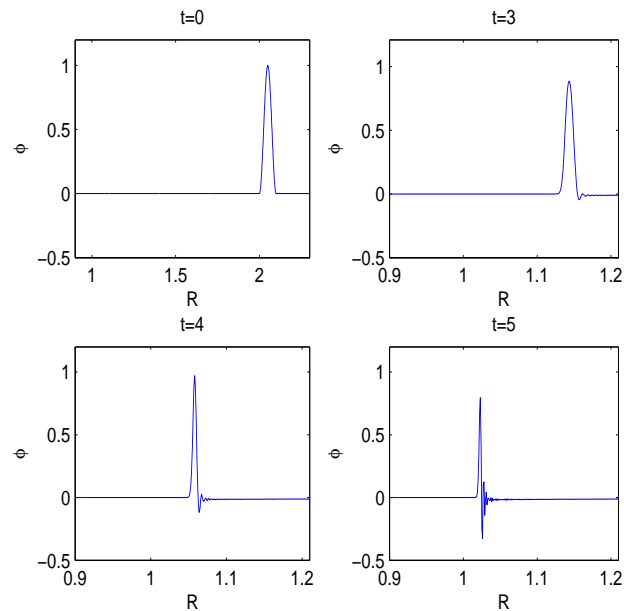}
\end{center}
  \caption{The figure shows four time steps in the propagation of a
  scalar pulse of characteristic length \( R \) towards a black hole. We see
  substantial deformation of the pulse on approaching $R=1$, accompanied
  by a propagation speed which tends to zero.  A video of this simulation
  can be found at \url{mendozza.org/sergio/bouncing-bh} and at
  \url{youtube.com/watch?v=jBoHQv2BrF0}. }
\end{figure}

 We note that the extensive literature on scattering of waves
from black holes treats the problem either from the point of
view of perturbations to the metric (non--vanishing field strength
for the wave), limit behaviour,  or using a Fourier decomposition analysis
\citep[e.g.][]{regge,vishveshwara,Staro73,Mashh74,chandra,malec00,malec01,malec97,nov,scattering,andersson,crispino,sasaki,castineiras},
or the recent article by Dolan \citep{dolan} and references therein.
That the full complexity of the problem makes a spectral decomposition
approach only an approximation, specially for $R \sim 1$, is already
suggested by the inconsistencies reached after the proposed $G(R)$
solution to equation (21). In equation (24), $K$ is seen to be dependent
on $R$, contradicting the original assumption.  We shall perform a
direct integration of the resulting equations with the aim of following
explicitly the full behaviour of the problem, in the vanishing
field strength limit, a test field interacting with the black hole.

In order to obtain numerical solutions to equation (\ref{DimLess}) we have
used an explicit finite-difference leapfrog scheme (central approximation
in both time and space) which affords a local truncation error (LTE)
of order $k^2 + h^2$, where $k$ is the fixed time step and $h$ is the
fixed position step. All of the results presented in this section have
been obtained using $k=h=0.001$, so that the LTE remains at all times
quite small. Note that in the numerical scheme, the right hand side of
equation(\ref {DimLess}) is evaluated, and used to obtain the evolution
of the field through the left hand side of equation(\ref{DimLess}). In
this way, there is no singularity at $R=1$, only a zero, which is in
any case explicitly avoided by choosing a discretisation which avoids
having a grid point at $R=1$.

It has also been checked that these choices of $k$ and $h$ satisfy the
well known Courant-Friedrichs-Lewy condition for the stability of an
explicit finite-difference scheme for a hyperbolic partial differential
equation, such as equation (\ref{DimLess}), \cite{ames}.  Reducing the
size of $k$ and $h$ by a factor of 10 yields results with only minimal
differences from those shown, and then, only at the peaks, proving
numerical convergence of the scheme.

\begin{figure}
\begin{center}
\includegraphics[height=8.5cm,width=9.3cm]{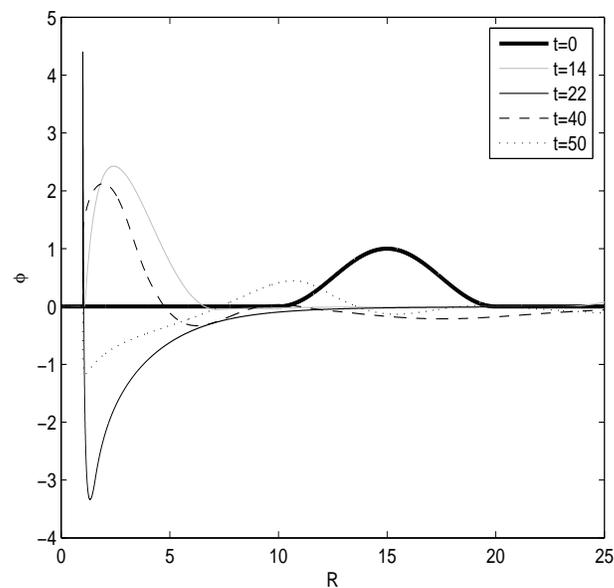}
\end{center}
  \caption{The figure shows four time steps in the propagation of a
  scalar pulse of characteristic length \( 10R \) towards a black hole. We see
  substantial deformation of the pulse as it approaches $R=1$, followed
  by the emission of a trainwave of wavelength comparable to the original
  wave \( 10R \), after a finite time. A video of the simulation can be
  obtained from \url{http://www.mendozza.org/sergio/bouncing-bh},
  \url{youtube.com/watch?v=u52CqjVerlQ} 
  and \url{youtube.com/watch?v=g1AruhdDzGA}.}
\end{figure}

Figure 1 shows results for a pulse of wavelength $0.1R$, a regime
where the standard small wavelength approximation would be expected to
hold. In the above, as in all that follows, the term wavelength refers to
the typical extent of the pulse when far from $ R= 1 $.  We see that
although propagation towards $R=1$ proceeds with deformations of
the original pulse, the pulse approaches $R=1$ with a speed given by
equation (\ref{Vel}) (see figure 4), which appears indistinguishable
from the $ds=0$ solution of $c=(R-1)/R$. The spatial extent of the pulse
is progressively reduced, in consistency with standard gravitational
blueshift. If played in reverse, we would see the gravitational redshift
of a wave emitted from an $R \rightarrow 1$, tending to infinity. The
pulse essentially stalls and would reach $R=1$ in an infinite time,
given the Schwarzschild description of the problem from the point of
view of a distant observer.

On the other hand, figure 2 shows the propagation of a pulse having
an initial extent of $10R$, starting at $R=10$. This pulse approaches
$R=1$, is deformed substantially, and produces a reflected wave train,
with a residual amplitude in the vicinity of $R=1$ which slowly decays.
We see a reflected pulse appearing in a finite coordinate time.
Notice that a wave solution ceases to be valid for large pulses,
as the original pulse shape is completely lost. This shows that any
description of the scattering problem in terms of a WKB treatment is
at best a first approximation as the evolution near \( R \sim 1 \)
invalidates the assumptions of spectral decomposition analysis, e.g. as
noted in \citep{sanchez01}, and treated in the improved analytical method
given there.  A single component in frequency space gives rise to complex
spectra as \( R \rightarrow 1 \) \citep{malec00,malec00b}.

In terms of the thought experiments of the opening section, we see that
a strictly standard mechanism naturally arises, such that scalar waves
larger than the Schwarzschild radius are essentially prevented from
entering the black hole.  It is interesting to see that the critical
wavelength, for the scalar field, appears approximately at precisely the
scale identified by the heuristic entropy considerations of Section~1,
when considering EM waves.  We have identified the critical wavelength to
lie somewhat below $1R$.  By comparing reflected scalar waves of various
initial wavelengths, we conclude that the characteristic wavelength
of the reflected wavelength is of the order of the initial wavelength,
as happens in the case of gravitational radiation \citep{vishveshwara}.

Away from $R=1$ we observe that, as expected, the physics is very similar
to spherical propagation, as the limiting form of equation (\ref{DimLess})
as $R \to \infty$ is precisely the equation for a spherically symmetric
scalar wave. As $R \to 1^+$, however, the propagation velocity decreases
noticeably and the wave amplitude increases, giving the impression
that the waveform is being smeared against $R=1$. This is explained
by the fact that the propagation term is proportional to  $(R-1)^2$
while the spherical spreading term is proportional to $(R-1)$, so that
the influence of the propagation term becomes smaller at a more rapid
rate than that of the spreading term.

\begin{figure}
\begin{center}
\includegraphics[height=8.5cm,width=9.3cm]{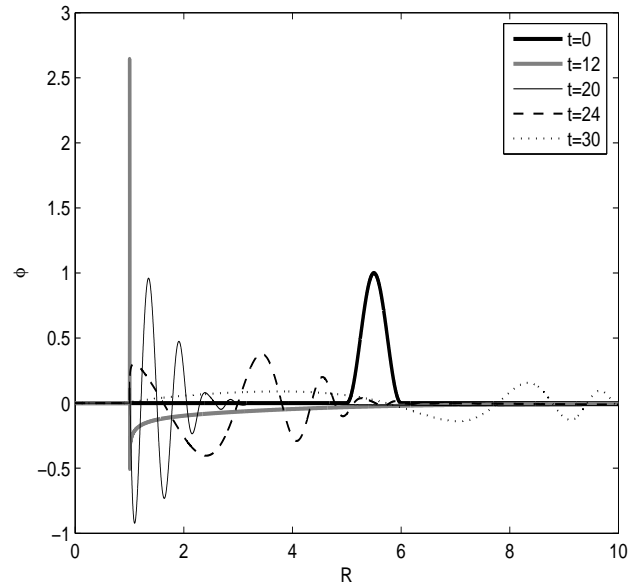}
\end{center}
  \caption{The figure shows four time steps in the propagation of a
  scalar pulse initially with an extent of $1R$
  towards a black hole. We see substantial deformation of the pulse
  as it approaches $R=1$, followed by the emission of a trainwave of
  wavelength comparable to the original wave, after a finite time.}
\end{figure}

Figure 3 is analogous to figure 2, but for a pulse having an initial
extent of $1R$, close to the critical value, we see an essentially
equivalent behaviour to that of the $10R$ pulse. The critical value for
the strong qualitative change in regime, from that of figure 1 to that of
figures 2 and 3, lying slightly below this point.

The preceding results are not unexpected, if one considers existing
studies of gravitational backscattering of light, where the fact that
light rays can travel along non-null geodesics is well known. We note
important precedents in the work of \citep{Staro73} who showed that in the
limit as $\lambda \rightarrow \infty$, electromagnetic and gravitational
waves will be reflected off Kerr black holes. Also, \citep{Mashh74}
demonstrated analytically, under the assumption that a strictly wave
solution should always apply, that the reflection coefficient for EM
waves tends to 1 for wavelengths of the order of $R_{S}$.  A comprehensive
study of scattering from black holes can be found in \citep{scattering}.
However a full numerical solution of scattering of a massless scalar
test field is not found in the literature, neither are the entropy
arguments relating limiting absorption wavelengths and the second law
of thermodynamics, as presented here.

More recent numerical studies include
\citep{malec00,malec00b,malec01,malec97} and references therein.  It has
been previously found that for the case of backscattering by Schwarzschild
black holes, although mostly treated in terms of perturbations on the
metric, the corresponding space--time works as a nonuniform medium with a
varying refraction index for electromagnetic waves. The magnitude of the
back-scattered wave depends on the frequency spectrum of the radiation:
it becomes negligible in the short wave limit and can be significant in
the long wave regime \citep{Kar, malec00, malec01}.  For scalar waves,
the full numerical treatment presented here suggests that this process
saturates and leads to zero absorption of light for wavelengths larger
than $R_{s}$.

Finally, we have calculated numerically the pulse propagation speeds.
We have evaluated propagation speeds as the time derivatives of the
position of the maximum of the pulse, they are hence phase velocities
in the case where the pulse retains its shape and behaves as a wave,
and group velocities when significant distortions in the shape of the
pulse appear.  Figure 4 shows a comparison of the actual pulse propagation
speeds, thick lines, and the approximation of equation(\ref{Vel}), thin
lines, as a function of radius, for the pulses shown in figures (1) and
(2). The pulse which started with a wavelength of $0.1R$ is shown in the
upper panel, and is seen to propagate exactly at the speed predicted by
equation (\ref{Vel}) for that wavelength. This was to be expected, as over
the extent of the pulse no significant variations in the metric occur,
until only very close to $R=1$. The wavelength of the pulse changes,
as seen in figure (1), but given the large initial value of $W=100$,
and the rapid approach to the asymptote of equation (\ref{Vel}) for
large $W$, this does not introduce significant variations.

\begin{figure}
\begin{center}
\includegraphics[height=9.0cm,width=9.3cm]{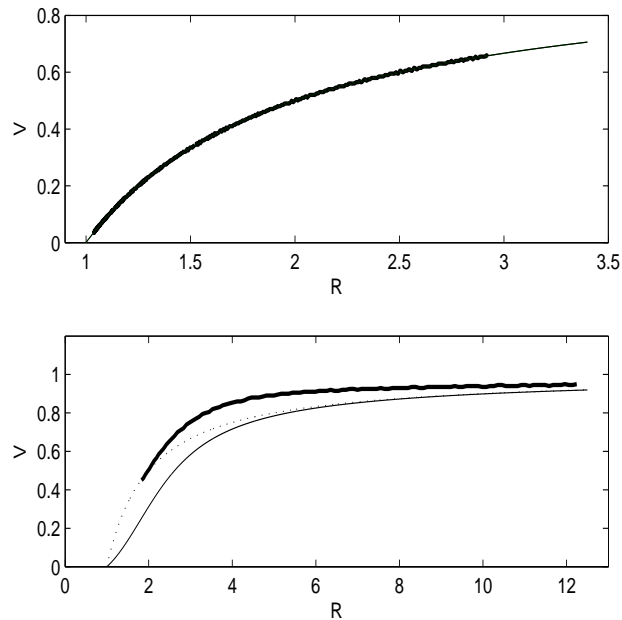}
\end{center}
  \caption{Propagation speeds $V$ of the maxima of the pulses shown
  in figures (1) and (2), upper and lower panels, respectively, thick
  lines, compared to the corresponding solutions of equation(\ref{Vel}),
  thin line. The dotted curves show the speed along a null geodesic,
  $c=(R-1)/R$.}
\end{figure}

The lower panel of figure (3) shows the propagation speed of a pulse
starting with a wavelength $10R$.  This time, the difference with
the speed predicted by equation (\ref{Vel}) is much more obvious, as
significant distortions in the metric, over the extent of the pulse,
are apparent inwards of around 1 wavelength, $R=10$.  In all cases, the
actual propagation speed is seen to deviate upwards of the prediction
of equation (\ref{Vel}), inwards of a certain critical radius. As in
the previous case, the asymptote towards $V=1$ is evident, both for the
numerical wave and the solutions of equation (\ref{Vel}). The function
$c=(R-1)/R$, the propagation corresponding to the standard vanishing
wavelength limit of $ds=0$, is also shown, seen completely superimposed
onto the other two curves for the short pulse, and appearing just below
the numerical speed in the $10R$ case. This last merely a result of the
pulse deformation, as the maximum moves within the pulse towards the
incoming edge of it, turning the numerical speed into a group velocity
of the problem.  The above highlighting the dispersive nature of the
problem and the intrinsic limitations of fully analytic approaches.
Results for the full electromagnetic vector field can be expected to be
qualitatively similar to the more limited scalar wave problem treated
here, e.g. \citep{andersson,dolan}.

  To check the validity of the numerical scheme, we have performed two
tests.  The first one regarding the sensibility of the problem to the
complications that may arise at \( R = 1 \), we have repeated all
simulations after shifting the numerical grid by a fraction of the radial
interval in such a way that the first grid point to the right of \( R = 1
\) changes slightly.  Results are completely insensitive to these shifts,
showing that the results are not sensitive to the coordinate singularity
that appears at \( R = 1 \) in the Schwarzschild space-time.  The
reflection we observe occurs very close to \( R = 1 \), still
outside the event horizon.  The second test has been an explicit
calculation of the total energy \( E \) of the scalar field, given by
\citep{Land}.

\begin{equation}
  E = 4 \pi \int{ T^{00} r^2 \mathrm{d} r },
\label{energy}
\end{equation}

\noindent where \( T^{00} \) is the time-time component of the stress-energy
tensor for a scalar field given by \citep{ryder} 

\begin{equation}
  T^{00} = g^{0\alpha} g^{0\beta}
\Psi,_\alpha \Psi,_\beta - g^{\mu\nu} \Psi,_\alpha \Psi,^\alpha / 2 =
\left( 1 / 2 \right) \left[ \left( 1 - 1/r \right)^{-1} \Psi,_t^2 +
\Psi,_r^2  \right].
\label{stress-energy}
\end{equation}

\noindent The net result of this calculation is that the total energy is
conserved in all simulations to better than one percent accuracy, over 100
units of T, \( 10^{5} \) simulation time steps.  Thus, the total initial
energy in the large pulse simulations remains the same when compared with
the energy after observing the wave far away from \( R = 1 \) after it has
``bounced'' completely. We have found full reflection of pulses larger
than the critical wavelength, to our numerical accuracy of 99\%. This
however, does not constitute a rigorous proof of full reflection for the
problem being treated, indeed, going back to the heuristic arguments of
section 1 on the EM field, the apparent inconsistency vanishes not only
in the case of full reflection of large wavelengths, but also for any
number of scenarios, provided large wavelengths are very substantially
reflected to possibly slightly varying degrees. Our numerical experiments
allow us to identify very substantial reflections of over 95\% of the
incoming energy in large pulses.  This is not seen to any degree in the
case of small pulses.

These results suggest the possibility of direct detection of black
holes through the study of radiation being reflected off them, as
already mentioned by \cite{fabbri,Kar}, in connection with fractional
backscattering in general. For stellar black holes, the gravitational
radii would be in the kilometre range, substantial EM backscattering
would hence be in this wavelength range, which is totally blocked by
the atmosphere, hence requiring presently non existent orbital or moon
based observatories for their detection. The possibility of much smaller
black holes appearing in particle accelerators has been suggested,
a case where substantial backscattering, would be expected to occur in
detectable ranges.

\section{Conclusions}

We have provided heuristic arguments suggesting that if a macroscopic
black hole is allowed to swallow classical Plank radiation colder than
itself, the Bekenstein interpretation of the horizon area of a black
hole as a measure of its entropy is not consistent with the second law of
thermodynamics.  In the macroscopic regime, entropy considerations suggest
a cut off limit for the light absorption spectrum of black holes, or very
substantial reflection of waves above a critical wavelength $\sim R$. It
is interesting to note that this critical wavelength is consistent with
previously identified critical wavelengths for the interaction of waves
and black holes, calculated in absence of any entropy considerations.

The standard $ds=0$ treatment of an eikonal approximation necessarily
fails when variations in the metric appear over a scale comparable
to the wavelengths present.  In the particular case of scalar waves,
this variations imply a non-achromatic effect for their interaction with
black holes, in particular, propagation velocities which fall below $c$,
even in the wave's proper frame, as the frequency is decreased.

Scalar waves having wavelengths longer than the Schwarzschild radius
of a black hole are very substantially reflected, or perhaps even fully
bounce off. It is interesting that the critical values appear precisely
at the wavelength scale identified through entropy considerations on
the interaction of Planck radiation and black holes.

\section{Acknowledgements}
The authors acknowledge comments made by Prof. A.~Starobinsky and Prof
B.~Carter.  We thank the comments made by three anonymous referees which
helped shape the last version of this paper.  X. Hernandez acknowledges
the support of grant UNAM DGAPA (IN117803-3), CONACyT (42809/A-1)
and CONACyT (42748).  C. Lopez-Monsalvo thanks economic support from
UNAM DGAPA (IN119203).  S. Mendoza gratefully acknowledges financial
support from CONACyT (41443) and UNAM DGAPA (IN119203). P. L. Rend\'on
acknowledges economic support from grant UNAM DGAPA (IN120008).


\begin{thebibliography}{99}

\bibitem{Bek82} J. D. Bekenstein, Phys. Rev, {\bf D 25}, 1527 (1982).

\bibitem{nov} V. P. Frolov \& I. D. Novikov, Black Hole Physics, Kluwer
   Academic Publishers, AH Dordrecht The Netherlands (1998).

\bibitem{unruh} W. G. Unruh, Phys. Rev. {\bf D 14}, 3251 (1976).

\bibitem{fabbri} R. Fabbri, Phys. Rev. {\bf D 12}, 933 (1975).

\bibitem{regge} T. Regge \& J.A. Wheeler, Phys. Rev. {\bf 108}, 1063   (1957).

\bibitem{vishveshwara} C.V. Vishveshwara, Nature {\bf 227}, 936 (1970).

\bibitem{scattering} J.A.H. Futterman, F.A. Handler \& R.A. Matzner,
Scattering from black holes, Cambridge Monographs on Mathematical Physics,
Cambridge University Press, 1st edition (1988).

\bibitem{Land} L. D. Landau \& E. M. Lifshitz, The Classical Theory of
Fields, Butterworth Heinemann, Oxford, 4rd edition (1995).


\bibitem{Tol} R. C. Tolman, Relativity, Thermodynamics and Cosmology,
  Oxford: Clarendon Press, (1934).

\bibitem{andersson} N. Andersson \& B. Jensen,  Scattering: Scattering and
Inverse Scattering in Pure and Applied Science, Eds. E.R. Pike \& P.C.
Sabatier, Academic Press (2001), arxiv:gr-qc/0011025.

\bibitem{crispino} L.C.B. Crispino, A. Higuchi \& G.E.A. Matsas, Phys. Rev.
    {\bf D 63} 7854 (1997).

\bibitem{wheeler} C. W. Misner, K. S. Thorne \& J. A. Wheeler,
Gravitation, Freeman, (1973).

\bibitem{friedrich}F. W. Hehl \& Y. N. Obukhov, Foundations of Classical
Electrodynamics: Charge, Flux, and Metric, Boston: Birkhauser, from the
series ``Progress in Mathematical Physics'', Volume 33 (2003).

\bibitem{dolan} S. R. Dolan arXiv:0801.3805v2 [gr-qc]

\bibitem{sasaki} M. Sassaki \& H. Tagoshi, Living Rev. Relativity {\bf 6},
   6 (2003).

\bibitem{castineiras} J. Casti\~neiras, L.C. Crispino, R. Murta \&  G.E.
Matsas, Phys. Rev. {\bf D 71}, 104013 (2005).

\bibitem{Kar}J. Karkowski, K. Roszkowski, Z. Swierczynski \& E. Malec,
Phys. Rev. {\bf D 67}, 064024 (2003).

\bibitem{ames}W. F. Ames, Numerical Methods for Partial Differential
 Equations, Thomas Nelson \& Sons, London (1969).

\bibitem{Staro73}A.A., Starobinsky \& S.M., Churilov, Sov. Phys. JETP
{\bf 38}, 1 (1973).

\bibitem{Mashh74}B. Mashhoon,  Phys. Rev. {\bf D 10}, 1059 (1974).

\bibitem{malec00}E. Malec,  Phys. Rev. {\bf D 62}, 084034 (2000).

\bibitem{sanchez01} N. Sanchez, Proceedings of Lectures delivered 
         at the Chalonge School, Nato ASI: Phase Transitions in the 
	 Early Universe: Theory and Observations. Editors H. J. de Vega, 
	 I. Khalatnikov,N. Sanchez. Kluwer Pub, hep-th/0106222, (2001)

\bibitem{malec00b}E. Malec,  Acta Phys.Polon. {\bf B 32}, 47 (2001).

\bibitem{malec01}J. Karkowski, E. Malec, Z. Swierczynski,
Class.Quant.Grav. {\bf 19}, 953-966 (2002).

\bibitem{malec97}E. Malec, arXiv:gr-qc/9711087 (1997).

\bibitem{chandra} S.~Chandrasekhar, ``The Mathematical Theory of Black
  Holes''. Oxford Classic Texts in the Physical Sciences, 
  Oxford University Press (2000).

\bibitem{ryder} L.H.~Ryder, ``Quantum Field Theory'', Cambridge University
Press (1996).

\end{thebibliography}

\end{document}